\documentclass[reprint,amsmath,amssymb,aps,pra,superscriptaddress]{revtex4-2}
\usepackage[utf8]{inputenc}
\usepackage{graphicx}
\usepackage[export]{adjustbox}
\usepackage{dcolumn}
\usepackage{bm}
\usepackage{float}
\usepackage{hyperref}
\usepackage{mathtools}
\usepackage{booktabs}
\usepackage{pgfplots}
\pgfplotsset{compat=newest}
\usepackage{algorithm}
\usepackage{algpseudocode}
\usepackage{colorspace}
\definecolorspace*{sRGB}{iccbased}{srgb.icc}

\newcolumntype{C}[1]{>{\centering\let\newline\\\arraybackslash\hspace{0pt}}m{#1}}

\newcommand*\from{\colon}

\newcommand*\Rr{\mathbb{R}}

\newcommand*\Zz{\mathbb{Z}}
\newcommand*\dif{\mathop{}\!\mathrm{d}}

\newcommand{\<}{\langle}
\let\>\undefined
\newcommand{\>}{\rangle}

\pgfplotsset{
  colormap/plasma/.style={%
    /pgfplots/colormap={plasma}{%
      rgb=(0.050383, 0.029803, 0.527975)
      rgb=(0.186213, 0.018803, 0.587228)
      rgb=(0.287076, 0.010855, 0.627295)
      rgb=(0.381047, 0.001814, 0.653068)
      rgb=(0.471457, 0.005678, 0.659897)
      rgb=(0.557243, 0.047331, 0.643443)
      rgb=(0.636008, 0.112092, 0.605205)
      rgb=(0.706178, 0.178437, 0.553657)
      rgb=(0.768090, 0.244817, 0.498465)
      rgb=(0.823132, 0.311261, 0.444806)
      rgb=(0.872303, 0.378774, 0.393355)
      rgb=(0.915471, 0.448807, 0.342890)
      rgb=(0.951344, 0.522850, 0.292275)
      rgb=(0.977856, 0.602051, 0.241387)
      rgb=(0.992541, 0.687030, 0.192170)
      rgb=(0.992505, 0.777967, 0.152855)
      rgb=(0.974443, 0.874622, 0.144061)
      rgb=(0.940015, 0.975158, 0.131326)
    },
  },
}

\begin{document}

\title{Robust self-assembly of nonconvex shapes in 2D}

\author{Lukas Mayrhofer}%
\email{lukas.mayrhofer@tum.de}
\affiliation{%
  Technische Universität München, Department of Mathematics, Boltzmannstraße 3, 85748 Garching, Germany
}%
\author{Myfanwy E. Evans}
\email{evans@uni-potsdam.de}
\affiliation{%
 University of Potsdam, Institute for Mathematics, Karl-Liebknecht-Str. 24-25, 14476 Potsdam, Germany
}%
\author{Gero Friesecke}
\email{friesecke@tum.de}
\affiliation{%
 Technische Universität München, Department of Mathematics, Boltzmannstraße 3, 85748 Garching, Germany
}%

\date{December 8, 2023}%

\begin{abstract}
We present fast simulation methods for the self-assembly of complex shapes in two dimensions.
The shapes are modeled via a general boundary curve and interact via a standard volume term promoting overlap and an interpenetration penalty.
To efficiently realize the Gibbs measure on the space of possible configurations we employ the hybrid Monte Carlo algorithm together with a careful use of signed distance functions for energy evaluation.

Motivated by the self-assembly of identical coat proteins of the tobacco mosaic virus which assemble into a helical shell, we design a particular nonconvex 2D model shape and demonstrate its robust self-assembly into a unique final state.
Our numerical experiments reveal two essential prerequisites for this self-assembly process: blocking and matching (i.e., local repulsion and attraction) of different parts of the boundary; and nonconvexity and handedness of the shape. 
\end{abstract}
\maketitle
\section{Introduction}
Virus capsids are formed by robust self-assembly from copies of a small number of different coat proteins into a unique, often highly symmetric configuration \cite{caspar_physical_1962}.
Yet the coat proteins themselves have nonsymmetric, nonconvex shapes which are not intuitive. An example is the single coat protein of the tobacco mosaic virus, whose copies self-assemble into a helical shell \cite{schmidli_tobacco_2019,schmidli_microfluidic_2019}. By contrast, the self-assembly of symmetric, convex shapes tends to lead to degenerate, non-unique configurations. Lennard-Jones clusters assembled from radially symmetric particles are prototypical \cite{wales_global_1997}; a detailed  analysis of the degeneracy for the simplified Heitmann-Radin potential in 2D is given in  
\cite{de_luca_classification_2017}.
The self-assembly of copies of a small number of convex shapes has also been systematically studied \cite{millan_self-assembly_2014}.

Here we investigate the self-assembly of identical copies of 2D shapes, and identify certain essential features which facilitate robust self-assembly into a unique configuration, among them nonconvexity. 

The shapes are modeled via a general boundary curve and their interaction is described by an energy functional which contains a standard volume term promoting overlap and an interpenetration penalty,  
\begin{equation}
  E := V + \gamma P.
\end{equation}
Here $\gamma > 0$ is a penalty strength. The penalty term $P$ is non-uniform along the boundary, to model repulsive and attractive regions. 

A common approach to simulate self-assembly is to use a particle model together with a Lennard-Jones pair-potential \cite{pakalidou_engineering_2020}.
In this study, we instead represent the shapes using signed distance functions which are well established in the computer graphics community \cite{perlin_hypertexture_1989,hart_sphere_1996}.
This type of representation is also used in the level-set method \cite{rautmann_finite_1980,osher_fronts_1988,osher_level_2003}, and allows us to efficiently evaluate the energy by numerical quadrature on a 2D grid.

For the evolution model, we use the hybrid Monte Carlo algorithm \cite{duane_hybrid_1987,betancourt_conceptual_2017} to compute approximate samples of the Gibbs measure
\begin{equation}
  \exp(-E(x) / T).
\end{equation}

\begin{figure}
    \centering
    \begin{adjustbox}{width=\columnwidth,center}
    \begin{tabular}{|c|c|c|c|}
        \hline
        \includegraphics[width=2cm,height=2cm,keepaspectratio,valign=m]{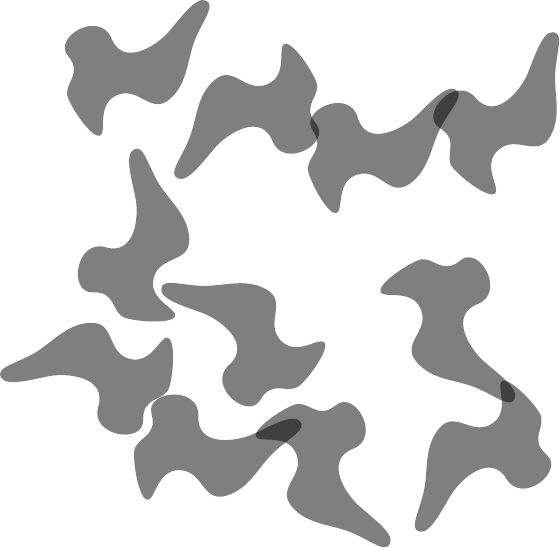} &
        \includegraphics[width=2cm,height=2cm,keepaspectratio,valign=m]{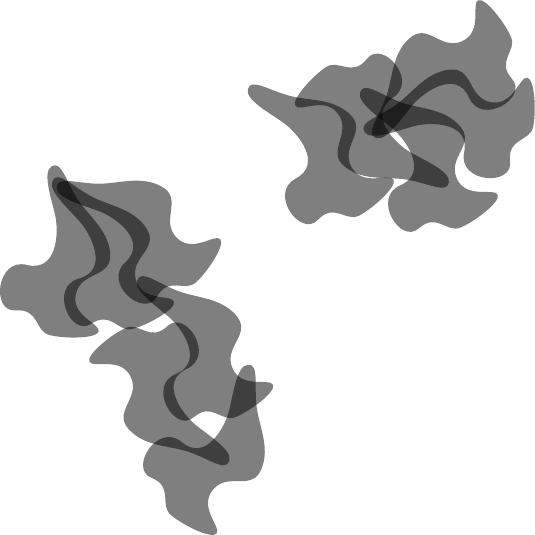} &
        \includegraphics[width=2cm,height=2cm,keepaspectratio,valign=m]{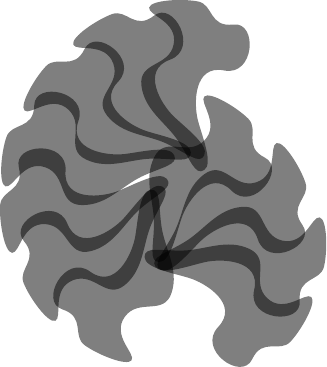} &
        \includegraphics[width=2cm,height=2cm,keepaspectratio,valign=m]{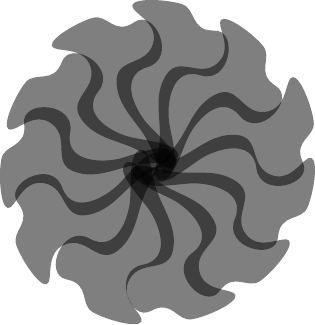} \\
        \hline
        \includegraphics[width=2cm,height=2cm,keepaspectratio,valign=m]{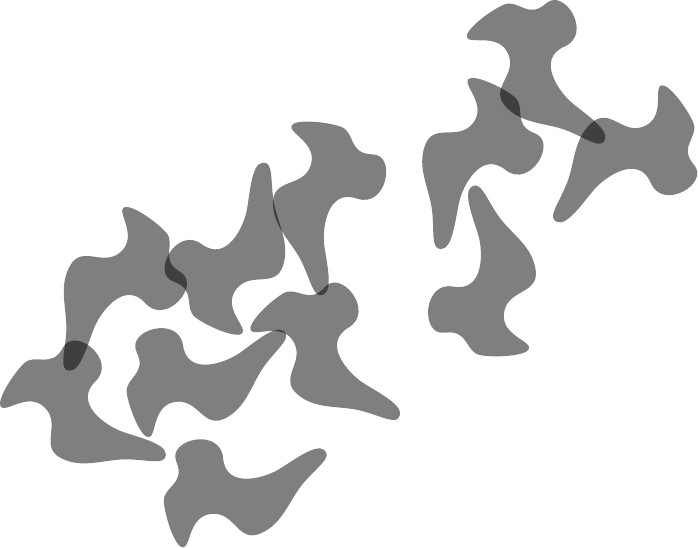} &
        \includegraphics[width=2cm,height=2cm,keepaspectratio,valign=m]{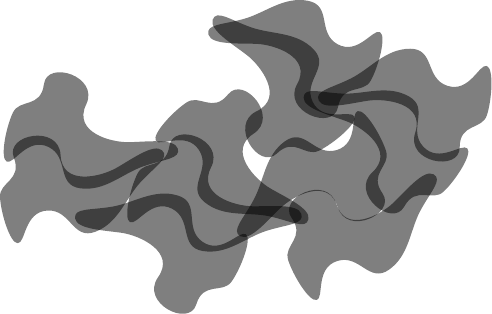} &
        \includegraphics[width=2cm,height=2cm,keepaspectratio,valign=m]{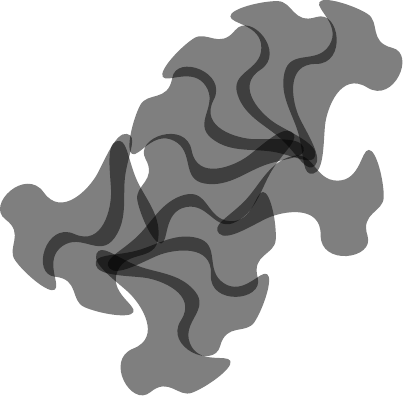} &
        \includegraphics[width=2cm,height=2cm,keepaspectratio,valign=m]{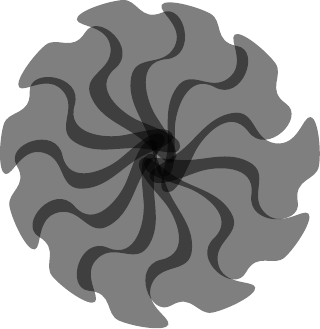} \\
        \hline
        \includegraphics[width=2cm,height=2cm,keepaspectratio,valign=m]{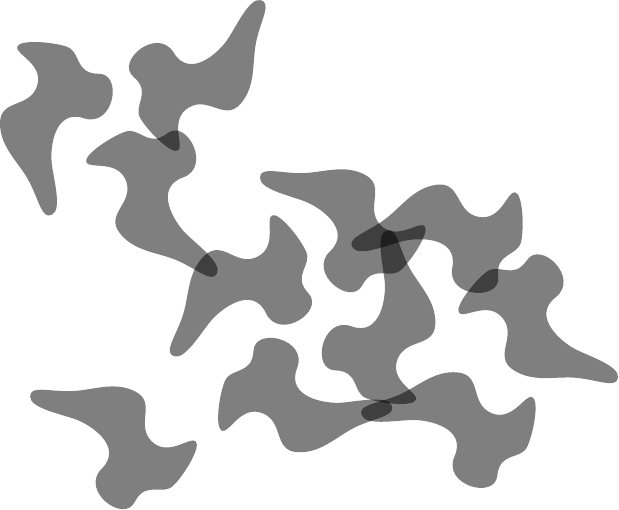} &
        \includegraphics[width=2cm,height=2cm,keepaspectratio,valign=m]{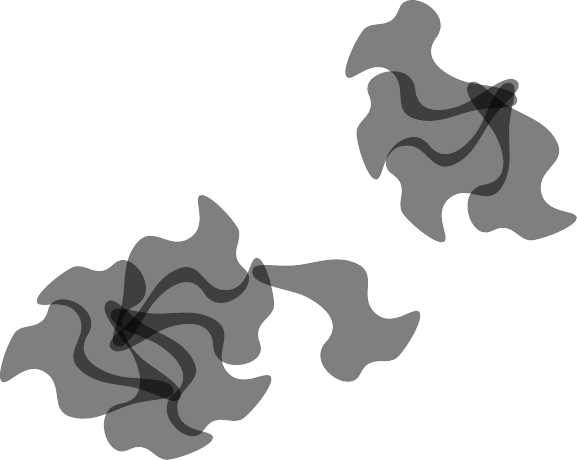} &
        \includegraphics[width=2cm,height=2cm,keepaspectratio,valign=m]{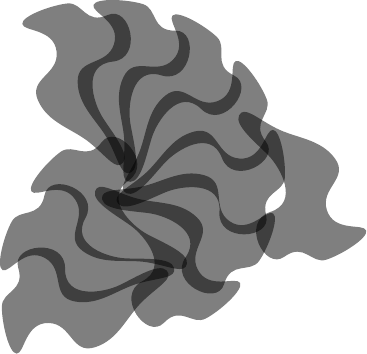} &
        \includegraphics[width=2cm,height=2cm,keepaspectratio,valign=m]{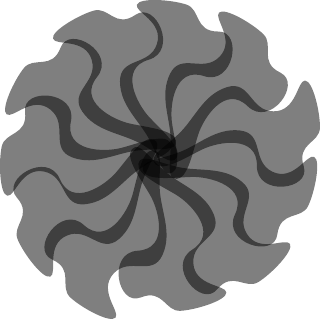} \\
        \hline
        initial states & $10^2$ iterations & $10^3$ iterations & $10^4$ iterations \\
        \hline
    \end{tabular}
    \end{adjustbox}
    \caption{\label{fig:shapestates}Assembly progress of our model shape during three simulation runs. Each row corresponds to a run. One can see that the model shape robustly self-assembles into a unique rotationally symmetric structure.}
\end{figure}

We design a nonconvex 2D model shape that achieves robust self-assembly into a unique structure, see Fig. \ref{fig:shapestates}.
The shape is inspired by the 3D shape of the TMV coat protein.
Numerical results show that self-assembly is sensitive to small changes in the shape, the energy, and the evolution model. In particular, we demonstrate how making the shape convex or making the penalty term uniform along the boundary curve leads to mis-assemblies. 

\section{\label{sec:energymodel}Energy model}
The total volume of $n$ given shapes $S_1, \dots, S_n \subset \Rr^2$ is
\begin{equation}
  \lambda\left(\bigcup_{k=1}^n S_k\right)
= \int_\Omega 1 - \prod_{k=1}^n (1 - \mathbf{1}_{S_k}(x)) \dif \lambda(x).
\end{equation}
We replace the indicator function $\mathbf{1}_{S_k}$ with a smooth approximation $H(\alpha f_k(\cdot))$, where $f_k$ is the signed distance function
\begin{equation}
    f_k(x) := \begin{cases}
        \operatorname{dist}(x, \partial S_k) & \text{if $x \in S_k$}, \\
        -\operatorname{dist}(x, \partial S_k) & \text{if $x \notin S_k$},
    \end{cases}
\end{equation}
$H$ is the smooth step function
\begin{equation}
    H(t) := \begin{cases}
        0 & \text{if $t < -1$}, \\
        \frac{3}{16} t^5-\frac{5}{8} t^3+\frac{15}{16} t+\frac{1}{2} & \text{if $-1 \leqslant t < 1$}, \\
        1 & \text{if $1 \leqslant t$},
    \end{cases}
\end{equation}
and $\alpha > 0$ is the transition speed.
This yields the volume term
\begin{equation}\label{eq:volumeterm}
\int_\Omega 1 - \prod_{k=1}^n (1 - H(\alpha f_k(x))) \dif \lambda(x).
\end{equation}

The overlap penalty term quantifies how much the shapes overlap.
Its role is to enforce an interpenetration constraint between the shapes so that the shapes do not collapse into trivial volume-minimizing structures in which all shapes overlap at a single point.

The strength of the overlap penalty can vary along the boundary of the shapes.
Some regions allow a lot of overlap, while other regions allow only minimal overlap.
We call these regions \emph{matching} and \emph{blocking} respectively.
To model matching and blocking, we associate a nonnegative penalty shift function $g_k \from \Rr^2 \to \Rr$ to each shape $S_k$ so that the larger $g_k$ is at a particular location in $S_k$, the more overlap is allowed at that location.

The overlap penalty term is given by
\begin{equation}\label{eq:penaltyterm}
    \int_\Omega \prod_{k=1}^n (1 + R(\beta [f_k(x) - g_k(x)])) \dif \lambda(x),
\end{equation}
where $R$ is the smooth ramp function
\begin{equation}
    R(t) := \begin{cases}
        0 & \text{if $t < -1$}, \\
        -\frac{1}{16} t^4+\frac{3}{8} t^2+\frac{1}{2} t+\frac{3}{16} & \text{if $-1 \leqslant t < 1$}, \\
        t & \text{if $1 \leqslant t$},
    \end{cases}
\end{equation}
and $\beta > 0$ is the steepness of the ramp.
Note that $g_k$ offsets the signed distance function to control how soon the penalty becomes active.

The volume \emph{interaction} term is the result after subtracting
\begin{equation}\label{eq:volumesubtract}
  \int_\Omega \sum_{k=1}^n H(\alpha f_k(x)) \dif \lambda(x)
\end{equation}
from the volume term and the penalty \emph{interaction} term is the result after subtracting
\begin{equation}\label{eq:penaltysubtract}
  \int_\Omega 1 + \sum_{k=1}^n R(\beta [f_k(x) - g_k(x)]) \dif \lambda(x).
\end{equation}
from the penalty term.
The subtracted terms are invariant under rigid movements applied to the shapes.

For our actual simulation, we put the shapes into the flat torus $\Omega := \Rr^2 / \eta \Zz^2$ of size $\eta > 0$.
The shapes $S_1, \dots, S_n$ are rigidly translated and rotated copies of a reference shape $S$.
Consequently, the signed distance functions $f_k$ and penalty shift functions $g_k$ are transformed versions of functions $f$ and $g$ associated to $S$.
The signed distance function $f$ is approximated using the procedure detailed in Appendix \ref{sec:sdfapprox}.

Integrals are evaluated using an equidistant quadrature grid with uniform weights.

\begin{figure}
\centering
\begin{tabular}{cccc}
    \includegraphics[width=.17\textwidth]{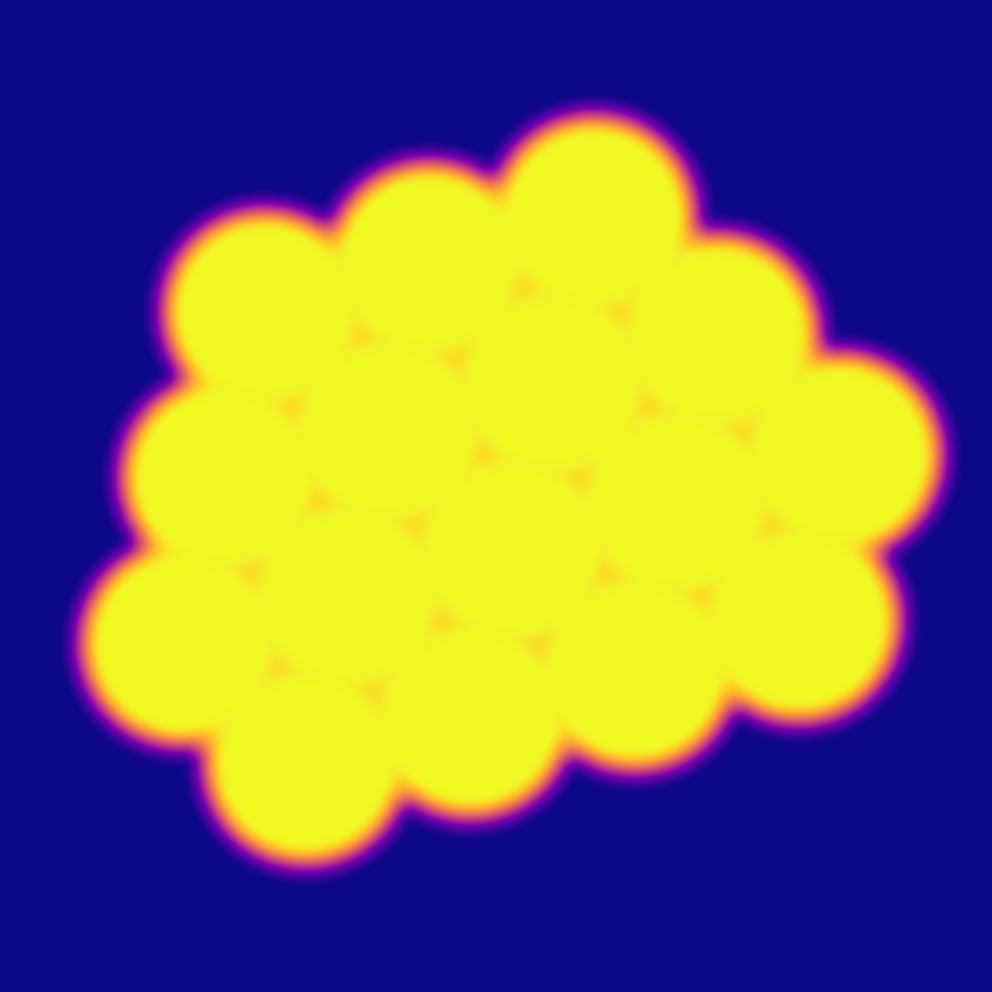} &
    \includegraphics[width=.17\textwidth]{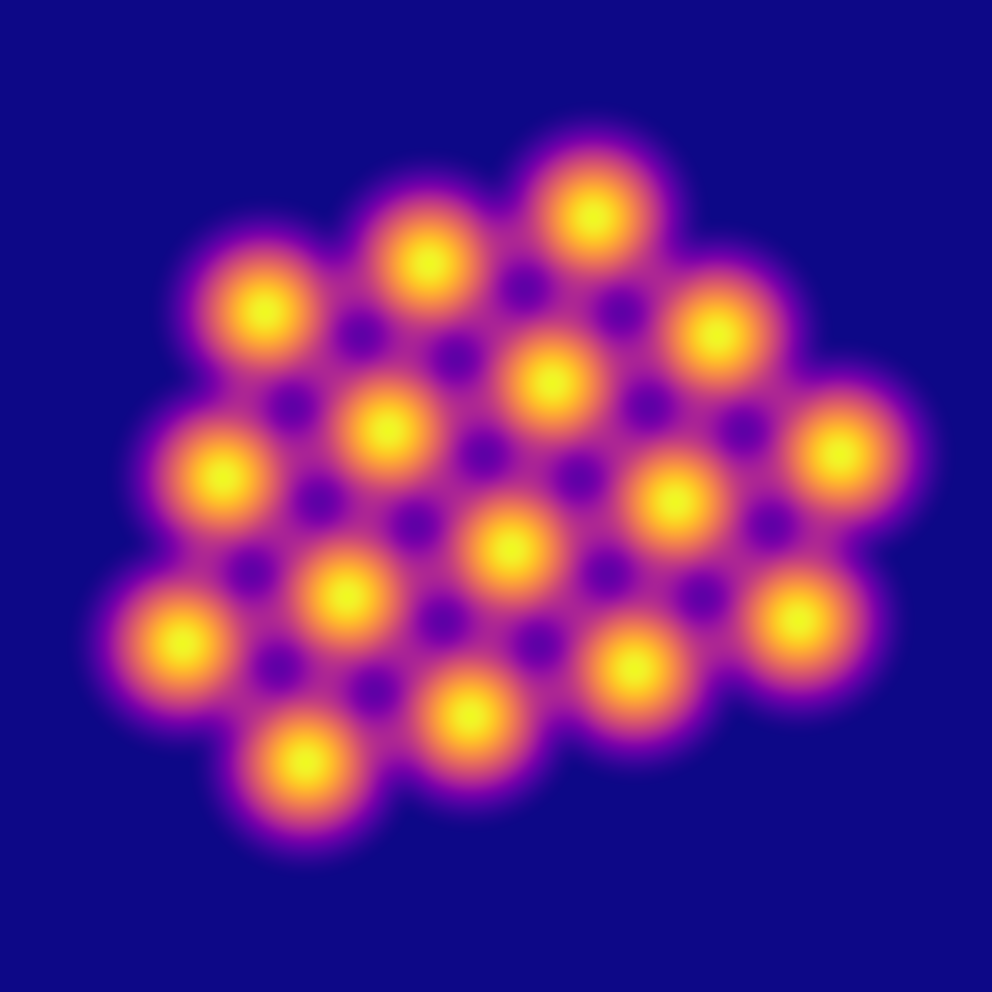} \\
    volume & penalty \\[0.7em]
    \includegraphics[width=.17\textwidth]{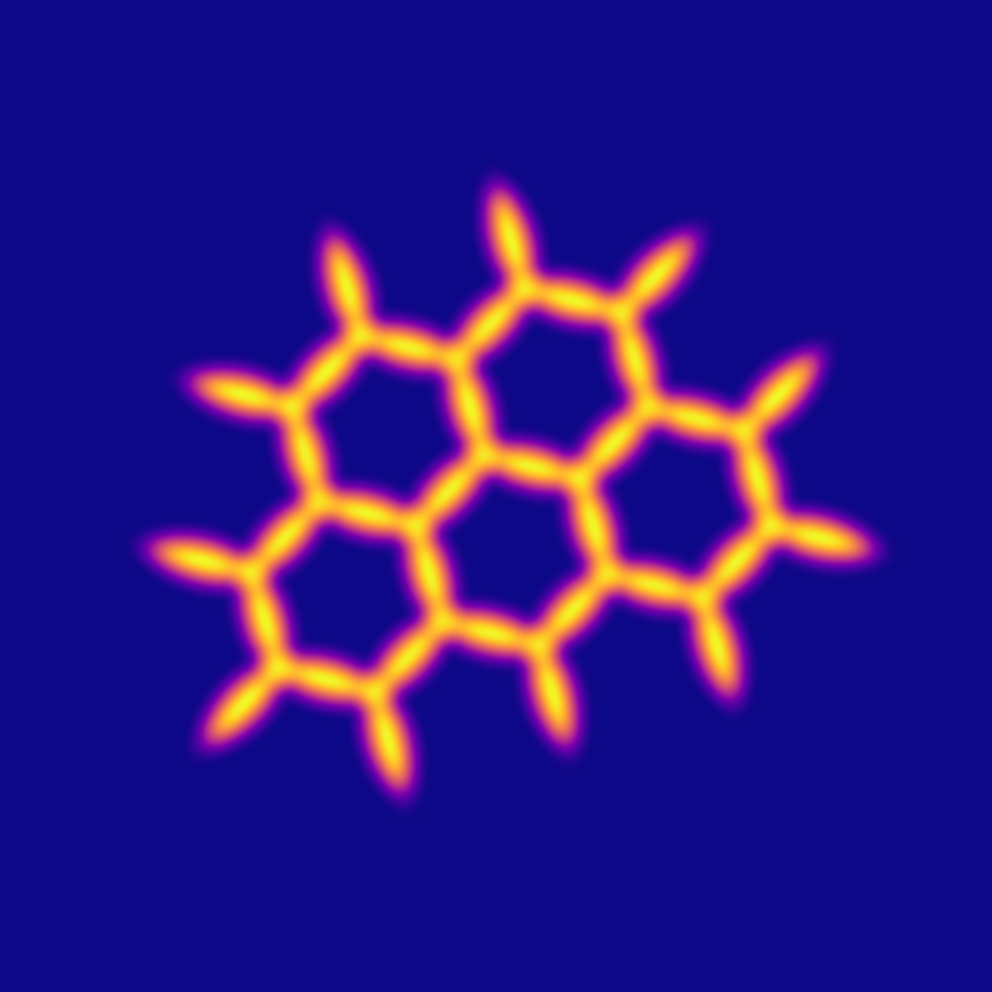} &
    \includegraphics[width=.17\textwidth]{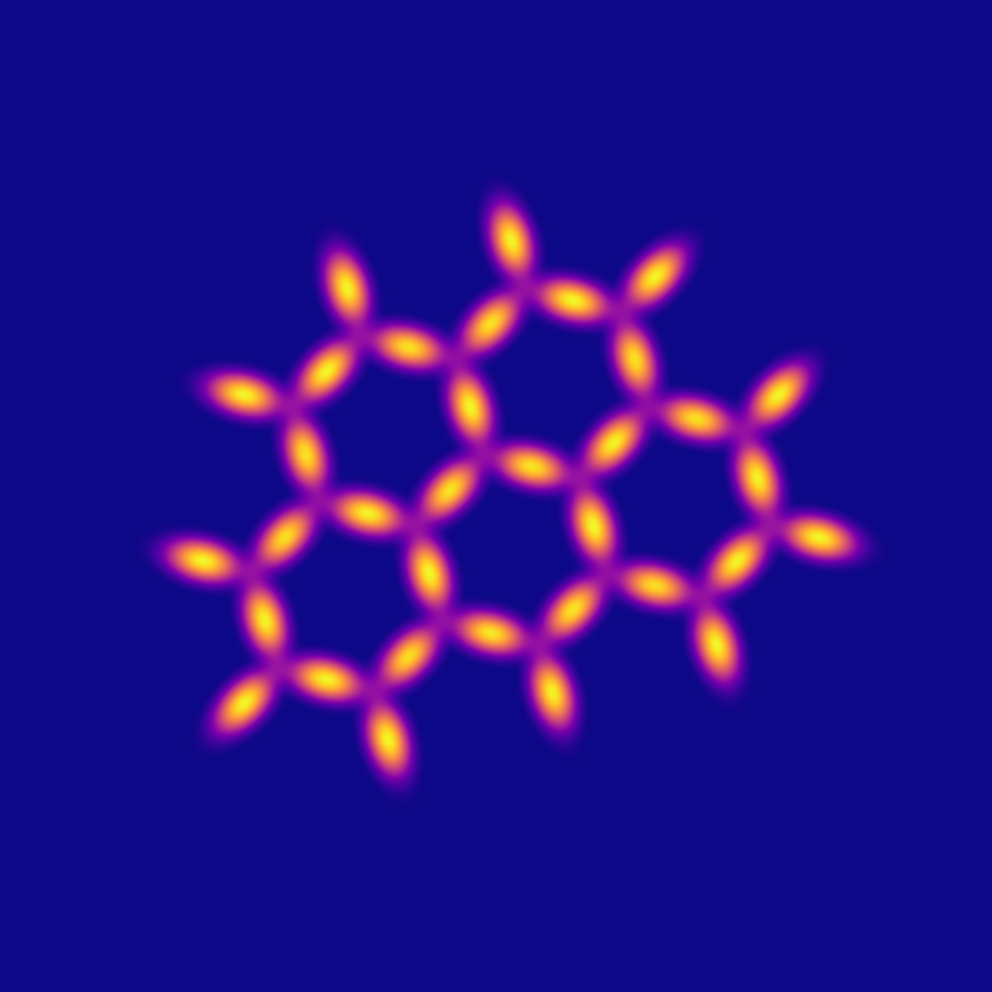} \\
    volume interaction & penalty interaction
\end{tabular}
\caption{\label{fig:ballintegrands}Visualization of the energy integrands: volume [eq.~\eqref{eq:volumeterm}], penalty [eq.~\eqref{eq:penaltyterm}], volume interaction [eq. \eqref{eq:volumeterm} minus eq.~\eqref{eq:volumesubtract}], penalty interaction [eq.~\eqref{eq:penaltyterm} minus 
eq.~\eqref{eq:penaltysubtract}]. For the volume interaction, whose integrand is negative, brighter colors denote lower values.}
\end{figure}

We visualize the energy terms for a configuration of $16$ balls in Fig. \ref{fig:ballintegrands}.
The penalty shift function is set to a positive constant for this visualization.

In summary, our energy model depends on the following parameters:
\begin{itemize}
    \setlength\itemsep{0em}
    \item shape ($S$)
    \item number of copies of the shape ($n$)
    \item size of ambient space ($\eta$)
    \item smooth step and ramp transition speeds ($\alpha, \beta$)
    \item penalty shift function ($g$) and penalty strength ($\gamma$)
    \item quadrature grid spacing
\end{itemize}

\section{\label{sec:evolutionmodel}Evolution model}
The $n$ copies of the shape can freely translate and rotate.
For each copy, the position and orientation is prescribed by a translation (an element of $\Rr^2 / \eta \Zz^2$) and a rotation (an element of $\mathrm{SO}(2)$) about the shape center.
This leads to the configuration space $G := (\Rr^2 / \eta \Zz^2 \times \mathrm{SO}(2))^n$, which is a compact Lie group.

We use the hybrid Monte Carlo (HMC) algorithm to compute samples from the Gibbs measure on $G$ given by
\begin{equation}
    \mu(A) := c \int_A \exp(-E(x)/T) \dif \lambda_G(x),
\end{equation}
where $c$ is normalization constant, $T > 0$ is a simulation temperature parameter, and $\lambda_G$ is the Haar measure on $G$.

An iteration of HMC consists of two steps:
First, a trajectory of the artificial Hamiltonian
\begin{equation}
    H(x, p) := E(x) / T + \frac12 \<p, p\>_\mathfrak{g}.
\end{equation}
is computed by integrating
\begin{align}
  x'(t) &= p(t) \\
  p'(t) &= (-1/T) \nabla E(x(t))
\end{align}
on the interval $[0, L]$ using the time-reversible and volume-preserving Leapfrog algorithm.
Here, $\<\cdot, \cdot\>_\mathfrak{g}$ is a fixed inner product on the Lie algebra $\mathfrak{g}$ corresponding to $G$ and $L > 0$ is a fixed trajectory length.
The initial value $x(0)$ is set to the current configuration of shapes and the initial velocity $p(0)$ is drawn from a Gaussian distribution with density proportional to
\begin{equation}
  \exp\left(-\frac12\<p, p\>_\mathfrak{g}\right).
\end{equation}
We employ the exponential map to adapt the Leapfrog algorithm to our Lie group setting.

The generated proposal $x(L)$ is accepted with probability
\begin{equation}
  \min \{ \exp(H(x(0), p(0)) - H(x(L), p(L))), 1 \}.
\end{equation}
If it is accepted, HMC sets the current state to $x(L)$ and generates a new proposal from this state.
Otherwise, a new proposal is generated from $x(0)$.

It can be shown that the Markov chain generated by this iteration leaves the Gibbs measure invariant.
Invariance is further discussed in Appendix \ref{sec:hmcinvariance}.

For the initial shape configuration, we draw a sample from the Haar measure on $G$, which means choosing initial translations uniformly in $[0, \eta)^2$ and rotations uniformly in $\mathrm{SO}(2)$.
To ensure that the shapes do not overlap before the actual simulation starts, we preprocess this sample using the Metropolis-Hastings algorithm with a simple repulsive energy model.

For our simulations, we select the inner product
\begin{equation}
    \<p, q\>_\mathfrak{g} := \sum_{i=1}^n \sigma_T^{-2}\<p_i^T, q_i^T\> + \sigma_R^{-2} p_i^R q_i^R,
\end{equation}
where $p_i^T, q_i^T$ are the translation components and $p_i^R, q_i^R$ the rotation components.
This means that the initial velocities for HMC are drawn from a normal distribution with a diagonal covariance matrix.
The parameters $\sigma_T$ and $\sigma_R$ define the standard deviations of the translation and rotation components respectively.

\section{\label{sec:results}Numerical results}

Our numerical experiments were carried out in the Julia programming language \cite{bezanson_julia_2017}.

\begin{figure}
  \centering
  \begin{adjustbox}{width=\columnwidth,center}
  \begin{tabular}{|C{2.5cm}|C{2.5cm}|C{2.5cm}|C{2.5cm}|}
    \hline
    \includegraphics[width=2.5cm,height=2.5cm,keepaspectratio,valign=m]{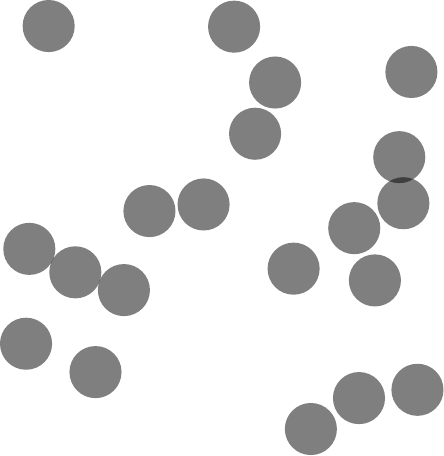} &
    \includegraphics[width=2.5cm,height=2.5cm,keepaspectratio,valign=m]{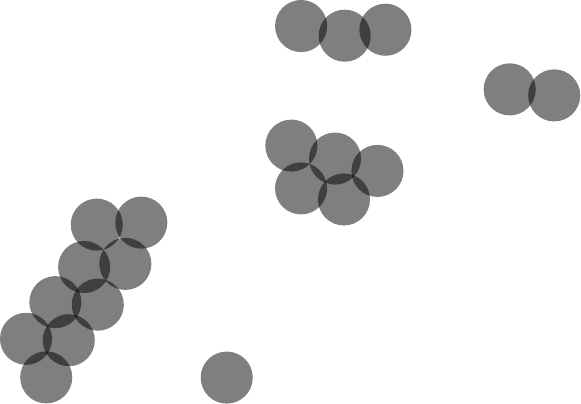} &
    \includegraphics[width=2.5cm,height=2.5cm,keepaspectratio,valign=m]{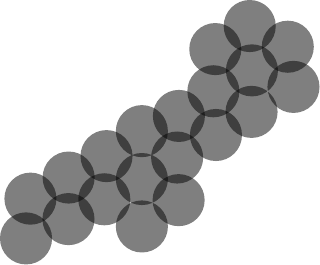} &
    \includegraphics[width=2.5cm,height=2.5cm,keepaspectratio,valign=m]{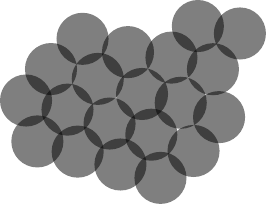} \\
    \hline
    \includegraphics[width=2.5cm,height=2.5cm,keepaspectratio,valign=m]{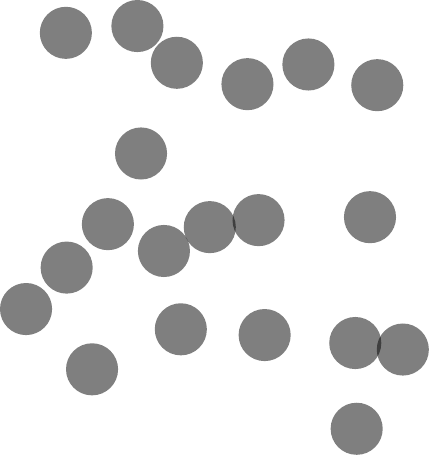} &
    \includegraphics[width=2.5cm,height=2.5cm,keepaspectratio,valign=m]{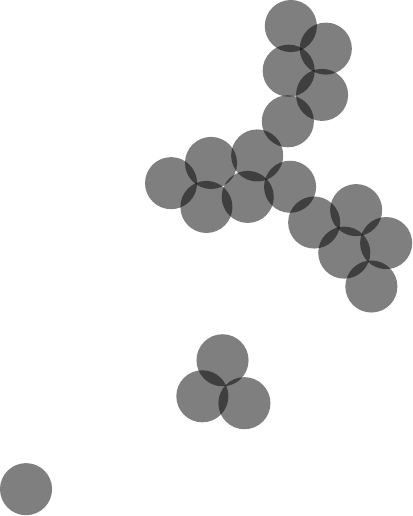} &
    \includegraphics[width=2.5cm,height=2.5cm,keepaspectratio,valign=m]{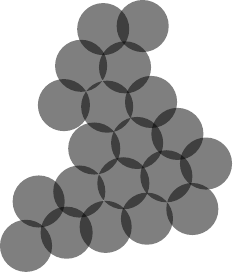} &
    \includegraphics[width=2.5cm,height=2.5cm,keepaspectratio,valign=m]{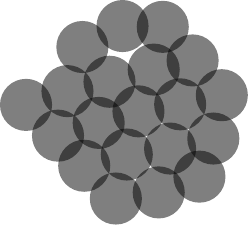} \\
    \hline
    \includegraphics[width=2.5cm,height=2.5cm,keepaspectratio,valign=m]{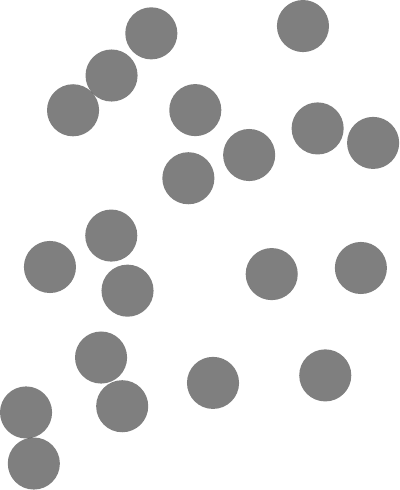} &
    \includegraphics[width=2.5cm,height=2.5cm,keepaspectratio,valign=m]{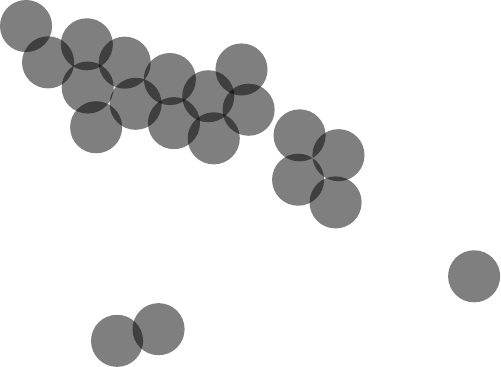} &
    \includegraphics[width=2.5cm,height=2.5cm,keepaspectratio,valign=m]{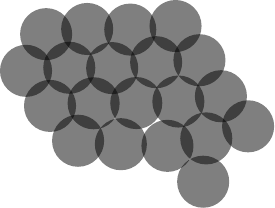} &
    \includegraphics[width=2.5cm,height=2.5cm,keepaspectratio,valign=m]{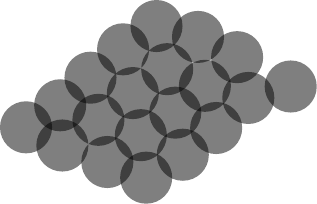} \\
    \hline
    initial states & $4 \cdot 10^1$ iterations & $4 \cdot 10^2$ iterations & $4 \cdot 10^3$ iterations \\
    \hline
  \end{tabular}
  \end{adjustbox}
  \caption{\label{fig:ballstates}Assembly progress of balls during three simulation runs. Each row corresponds to a run. One can see that the balls assemble into different configurations.}
\end{figure}

Simulations with simple convex shapes such as balls do not settle into a unique state, see Fig. \ref{fig:ballstates}.

\begin{figure}
    \centering
    \includegraphics[width=0.25\textwidth]{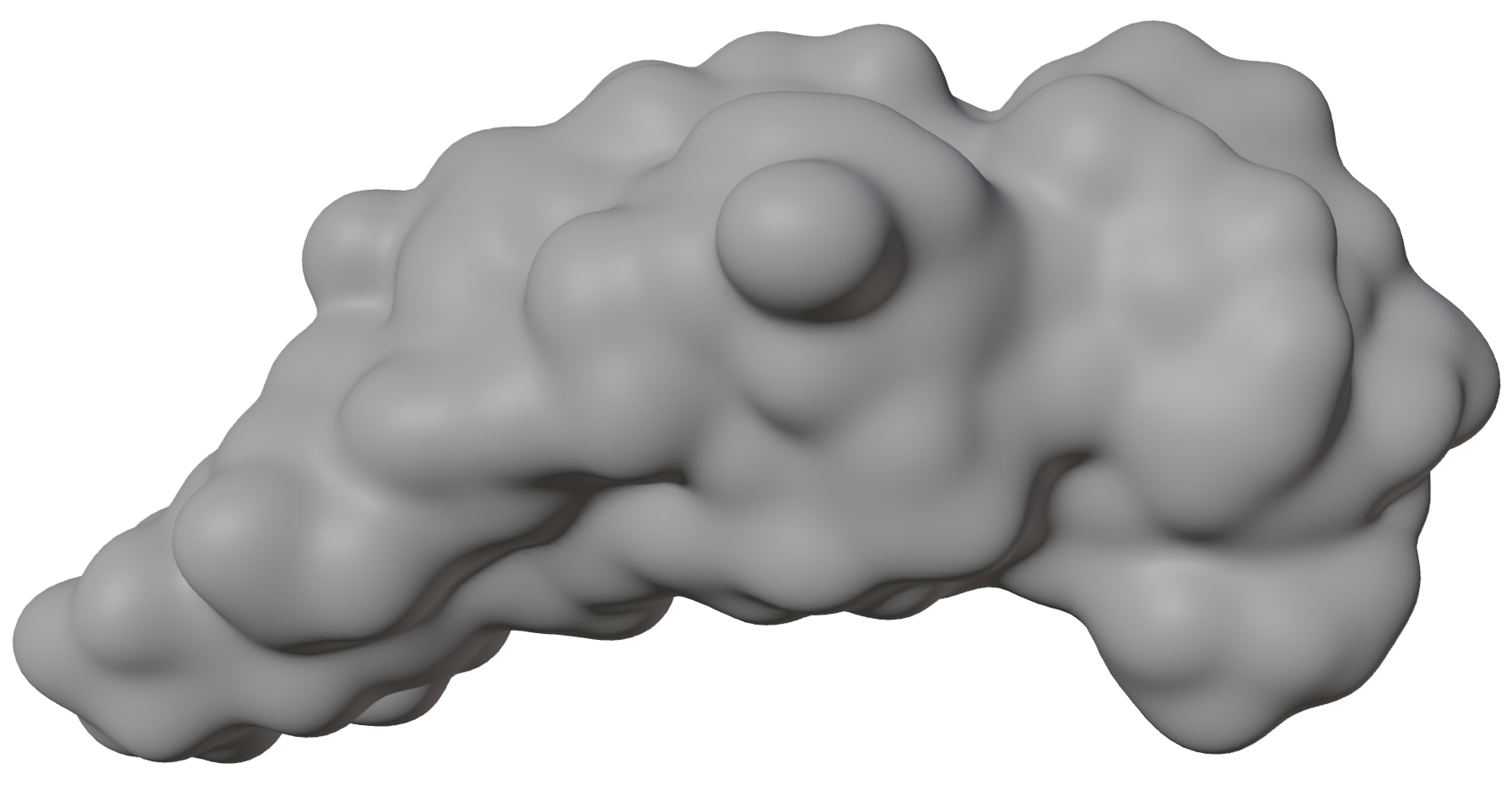}
    \caption{\label{fig:tmv}Coat protein of the tobacco mosaic virus \cite{schmidli_tobacco_2019,schmidli_microfluidic_2019}}
    \includegraphics[width=0.25\textwidth]{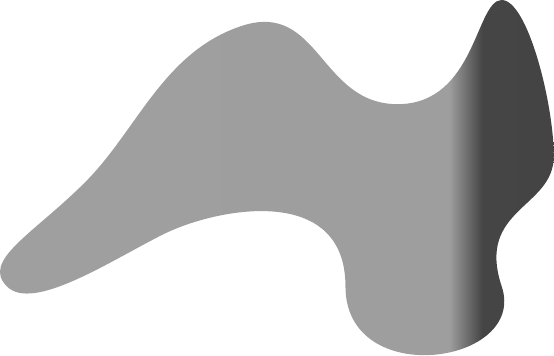}
    \caption{\label{fig:shapeparts}Model protein shape. Like the 3D shape in Fig.~\ref{fig:tmv} it is asymmetric, nonconvex and handed. The shape is subdivided into a light \emph{matching} part and dark \emph{blocking} part.}
\end{figure}

The coat protein of the tobacco mosaic virus (see Fig. \ref{fig:tmv}) was designed by evolution to robustly assemble into a helix.
We design a 2D model shape (see Fig. \ref{fig:shapeparts}) that shares some important features with the coat protein: it is highly asymmetric, nonconvex, and handed.
We want to investigate how these features affect the self-assembly of the model shape.

The penalty shift function is set to $0.1$ on the blocking part and to $0.5$ on the matching part with a smooth transition between these values.
A detailed description of the shape is given in Appendix \ref{sec:shapedefinition}.

We conduct assembly experiments with the simulation parameters in Table \ref{tab:shapepars}.
Using these parameters, we observe robust self-assembly into a unique configuration (see Fig. \ref{fig:shapestates}).

To quantify the self-assembly progress, we consider the graph of correctly attached shapes.
In this assembly graph, the vertices represent the shapes and the edges represent connections between shapes that are (within a small tolerance in relative translation and rotation) correctly attached to each other.
The number of connected components in this graph is a useful measure of self-assembly progress.

Assembly graphs with a single connected component closely correspond to correctly assembled configurations.
Therefore, we use this as our assembly criterion in the upcoming numerical experiments.

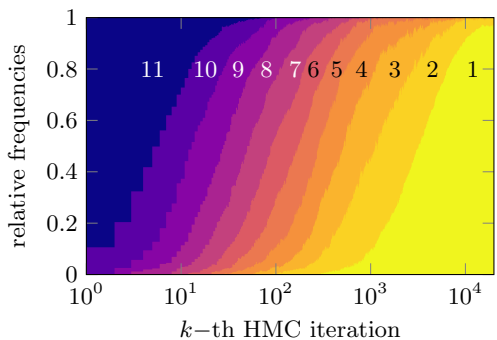
\begin{figure}
    \centering
    \begin{tikzpicture}
\begin{axis}[
xlabel={\small $k-$th HMC iteration},
ylabel={\small relative frequencies},
width={7cm}, height={5cm},
xmode={log}, xmin=1, xmax=20000, ymin=0, ymax=1, xminorticks=false, axis on top]
\addplot graphics [
    xmin = 1,
    xmax = 20000,
    ymin = 0,
    ymax = 1,
] {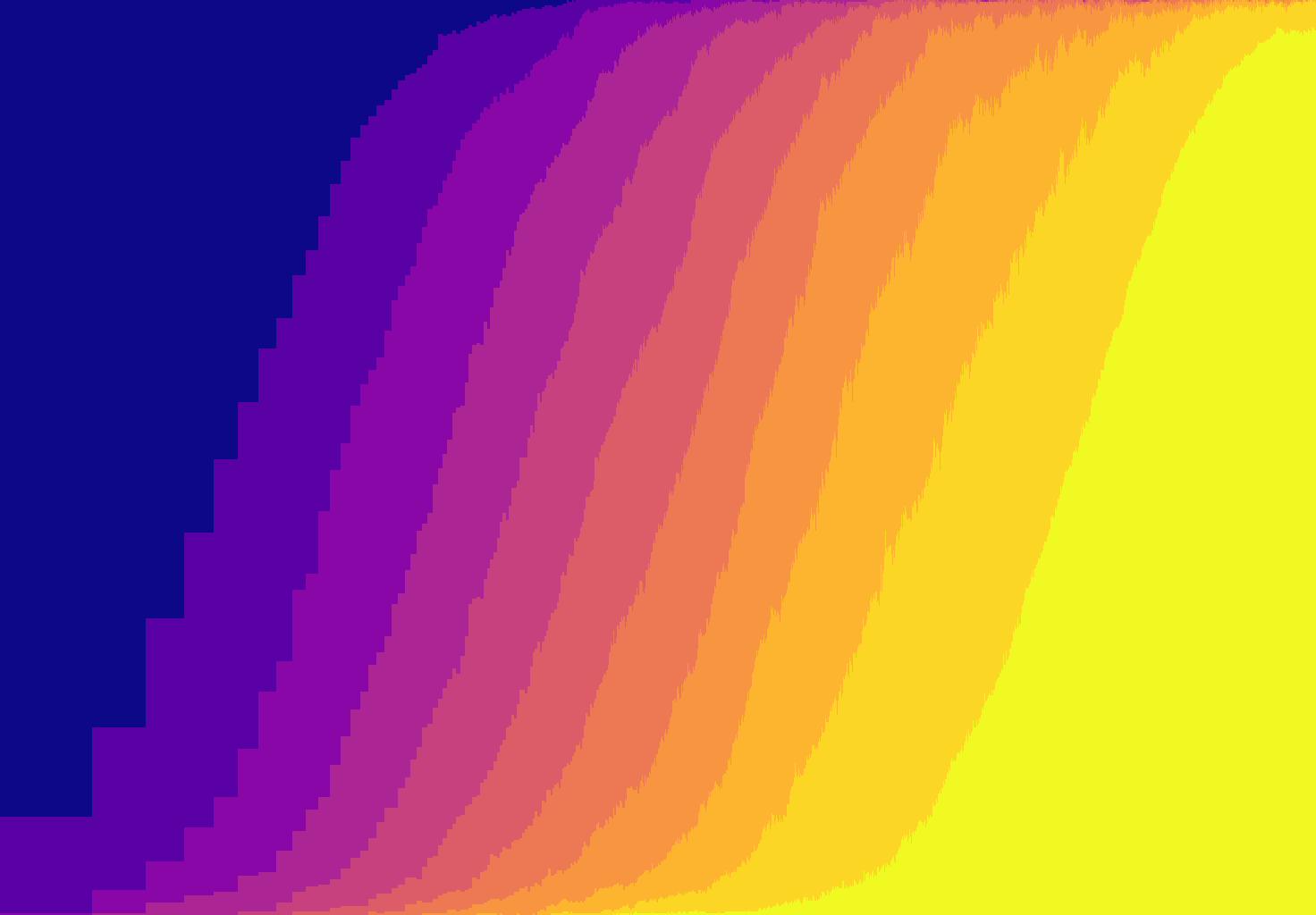};

\node[black] at (axis cs:12000, 0.8) {1}; %
\node[black] at (axis cs:4500, 0.8) {2}; %
\node[black] at (axis cs:1800, 0.8) {3}; %
\node[black] at (axis cs:800, 0.8) {4}; %
\node[black] at (axis cs:440, 0.8) {5}; %
\node[black] at (axis cs:250, 0.8) {6}; %
\node[white] at (axis cs:160, 0.8) {7}; %
\node[white] at (axis cs:80, 0.8) {8}; %
\node[white] at (axis cs:40, 0.8) {9}; %
\node[white] at (axis cs:18, 0.8) {10}; %
\node[white] at (axis cs:5, 0.8) {11}; %
\end{axis}
\end{tikzpicture}
    \caption{
    Assembly progress from $11$ connected components (disassembled configuration) to $1$ connected component (fully assembled configuration).
    The distribution over $512$ hybrid Monte Carlo runs is shown.
    }
    \label{fig:shapeconncomp}
\end{figure}

In Fig. \ref{fig:shapeconncomp} we show the distribution of the number of connected components as a function of HMC iterations.
Initially, there are typically $11$ connected components, which gradually reduce to $1$.
At $10^4$ iterations, most runs are in a correctly assembled configuration.

\subsection*{Parameter identification}
To achieve rapid and robust self-assembly, we need
\begin{itemize}
    \item \textbf{Convergence.} The evolution algorithm should converge quickly to the Gibbs measure.
    \item \textbf{The right Gibbs measure.} Samples drawn from the Gibbs measure should be in the assembled state with high probability.
\end{itemize}

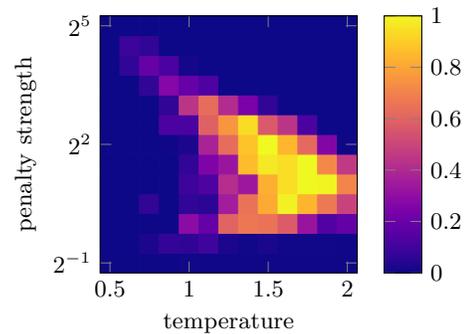
\begin{figure}
    \begin{center}
    \begin{tikzpicture}
\begin{axis}[%
  width={5.0cm},
  height={5.0cm},
  enlargelimits=false,
  xlabel = {temperature},
  ylabel = {penalty strength},
  ymode = {log},
  log basis y={2},
  colorbar,
  colormap/plasma,
  colorbar style={%
  },
  point meta min=0,
  point meta max=1,
  axis on top,
]

\addplot[
matrix plot*,
mesh/cols=13,
point meta=explicit,
]
table[meta=C]{
x y C
0.5 0.5 0.0011875
0.625 0.5 0.0
0.75 0.5 0.0015625
0.875 0.5 0.002
1.0 0.5 0.00725
1.125 0.5 0.003375
1.25 0.5 0.006125
1.375 0.5 0.0053125
1.5 0.5 0.0029375
1.625 0.5 0.0055
1.75 0.5 0.00575
1.875 0.5 0.0061875
2.0 0.5 0.0048125
0.5 0.7071067811865476 0.0
0.625 0.7071067811865476 0.0
0.75 0.7071067811865476 0.028
0.875 0.7071067811865476 0.0700625
1.0 0.7071067811865476 0.0601875
1.125 0.7071067811865476 0.095
1.25 0.7071067811865476 0.0265625
1.375 0.7071067811865476 0.0063125
1.5 0.7071067811865476 0.0225
1.625 0.7071067811865476 0.007875
1.75 0.7071067811865476 0.0095625
1.875 0.7071067811865476 0.008
2.0 0.7071067811865476 0.00475
0.5 1.0 0.0
0.625 1.0 0.0
0.75 1.0 0.0021875
0.875 1.0 0.000125
1.0 1.0 0.286125
1.125 1.0 0.183875
1.25 1.0 0.5990625
1.375 1.0 0.6618125
1.5 1.0 0.6479375
1.625 1.0 0.5731875
1.75 1.0 0.3238125
1.875 1.0 0.1404375
2.0 1.0 0.180625
0.5 1.4142135623730951 0.0
0.625 1.4142135623730951 0.0
0.75 1.4142135623730951 0.0625
0.875 1.4142135623730951 0.0
1.0 1.4142135623730951 0.0625625
1.125 1.4142135623730951 0.0509375
1.25 1.4142135623730951 0.351
1.375 1.4142135623730951 0.6491875
1.5 1.4142135623730951 0.809625
1.625 1.4142135623730951 0.9815
1.75 1.4142135623730951 0.835875
1.875 1.4142135623730951 0.6636875
2.0 1.4142135623730951 0.48675
0.5 2.0 0.0
0.625 2.0 0.0
0.75 2.0 0.0
0.875 2.0 0.0
1.0 2.0 0.0883125
1.125 2.0 0.127375
1.25 2.0 0.41375
1.375 2.0 0.3335625
1.5 2.0 0.82225
1.625 2.0 0.939125
1.75 2.0 0.999375
1.875 2.0 0.9884375
2.0 2.0 0.70125
0.5 2.8284271247461903 0.0
0.625 2.8284271247461903 0.0
0.75 2.8284271247461903 0.0
0.875 2.8284271247461903 0.0
1.0 2.8284271247461903 0.0625
1.125 2.8284271247461903 0.2520625
1.25 2.8284271247461903 0.325
1.375 2.8284271247461903 0.8549375
1.5 2.8284271247461903 0.963
1.625 2.8284271247461903 0.939
1.75 2.8284271247461903 0.9985
1.875 2.8284271247461903 0.714625
2.0 2.8284271247461903 0.451875
0.5 4.0 0.0
0.625 4.0 0.0
0.75 4.0 0.0
0.875 4.0 0.004875
1.0 4.0 0.027625
1.125 4.0 0.376125
1.25 4.0 0.62675
1.375 4.0 0.875
1.5 4.0 1.0
1.625 4.0 0.8405
1.75 4.0 0.6168125
1.875 4.0 0.2629375
2.0 4.0 0.0055
0.5 5.656854249492381 0.0
0.625 5.656854249492381 0.0
0.75 5.656854249492381 0.0
0.875 5.656854249492381 0.125
1.0 5.656854249492381 0.1206875
1.125 5.656854249492381 0.5903125
1.25 5.656854249492381 0.7505
1.375 5.656854249492381 0.9375625
1.5 5.656854249492381 0.5648125
1.625 5.656854249492381 0.438875
1.75 5.656854249492381 0.1248125
1.875 5.656854249492381 0.0
2.0 5.656854249492381 0.0
0.5 8.0 0.0
0.625 8.0 0.0
0.75 8.0 0.0
0.875 8.0 0.0625
1.0 8.0 0.4378125
1.125 8.0 0.62625
1.25 8.0 0.404875
1.375 8.0 0.2400625
1.5 8.0 0.0624375
1.625 8.0 0.0
1.75 8.0 0.0
1.875 8.0 0.0
2.0 8.0 0.0
0.5 11.313708498984761 0.0
0.625 11.313708498984761 0.0
0.75 11.313708498984761 0.0625
0.875 11.313708498984761 0.2671875
1.0 11.313708498984761 0.1875
1.125 11.313708498984761 0.1329375
1.25 11.313708498984761 0.000125
1.375 11.313708498984761 0.0
1.5 11.313708498984761 0.0
1.625 11.313708498984761 0.0
1.75 11.313708498984761 0.0
1.875 11.313708498984761 0.0
2.0 11.313708498984761 0.0
0.5 16.0 0.0
0.625 16.0 0.0625
0.75 16.0 0.1875
0.875 16.0 0.125
1.0 16.0 0.0120625
1.125 16.0 0.0
1.25 16.0 0.0
1.375 16.0 0.0
1.5 16.0 0.0
1.625 16.0 0.0
1.75 16.0 0.0
1.875 16.0 0.0
2.0 16.0 0.0
0.5 22.627416997969522 0.0
0.625 22.627416997969522 0.0864375
0.75 22.627416997969522 0.06275
0.875 22.627416997969522 0.0
1.0 22.627416997969522 0.0
1.125 22.627416997969522 0.0
1.25 22.627416997969522 0.0
1.375 22.627416997969522 0.0
1.5 22.627416997969522 0.0
1.625 22.627416997969522 0.0
1.75 22.627416997969522 0.0
1.875 22.627416997969522 0.0
2.0 22.627416997969522 0.0
0.5 32.0 0.0
0.625 32.0 0.0
0.75 32.0 0.0
0.875 32.0 0.0
1.0 32.0 0.0
1.125 32.0 0.0
1.25 32.0 0.0
1.375 32.0 0.0
1.5 32.0 0.0
1.625 32.0 0.0
1.75 32.0 0.0
1.875 32.0 0.0
2.0 32.0 0.0
};
\end{axis}
\end{tikzpicture}
    \caption{\label{fig:phaseplot_asmprob}%
    Self-assembly robustness.
    Shows the proportion of configurations that are correctly assembled.
    The proportions are computed from iterations $9 \cdot 10^3$ to $10^4$ of $16$ hybrid Monte Carlo runs for each choice of parameters.}
    \end{center}
\end{figure}

We investigate how different choices of simulation temperature and penalty strength affect the robustness of self-assembly.
For each choice, we tune the Leapfrog stepsize to achieve an acceptance rate of about $65\%$.
The results are shown in Fig. \ref{fig:phaseplot_asmprob}.

It turns out that there is a sweet spot for both parameters:
Low temperatures give more weight to low energy states but slow down the convergence to the Gibbs measure.
High temperatures converge faster but make the interesting configurations less likely to occur.
Low penalty strengths give the evolution algorithm more leeway to perturb shape configurations but increase shape overlap and the probability of the shapes collapsing.
High penalty strengths make the configurations unstable and likely to disassociate.

\subsection*{Ablation study}
\begin{figure}
    \centering
    \begin{tabular}{|C{2cm}|}
        \hline
        \includegraphics[width=2cm,height=2cm,keepaspectratio,valign=m]{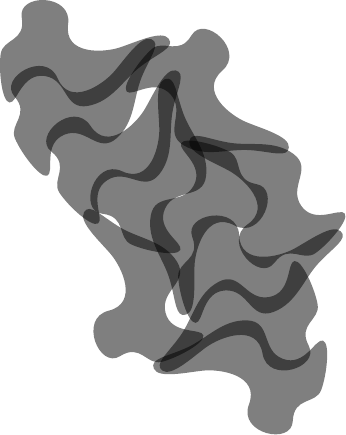} \\
        \hline
        \includegraphics[width=2cm,height=2cm,keepaspectratio,valign=m]{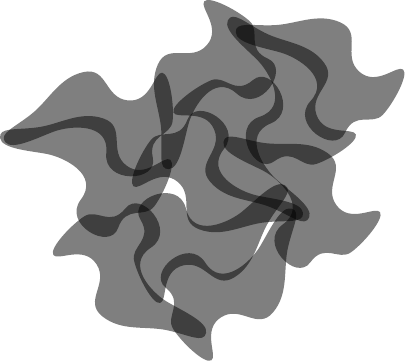} \\
        \hline
    \end{tabular}
    \hspace{0.3cm}
    \begin{tikzpicture}[baseline={(0,0.9cm)}]
\begin{axis}[xlabel={penalty shift on blocking part},ylabel={self-assembly rate}, width={5cm}, height={4cm}, grid={major}, xmin=0, xmax=0.5, ymin=0, ymax=1, xtick={0.0,0.1,0.2,0.3,0.4,0.5}]
    \addplot[color=red,mark=*]
        coordinates {
            (0.0,0.87809375)
            (0.03333333333333333,0.96875)
            (0.06666666666666667,0.9085625)
            (0.1,0.9538125)
            (0.13333333333333333,0.85803125)
            (0.16666666666666666,0.8979375)
            (0.2,0.76953125)
            (0.23333333333333334,0.7154375)
            (0.26666666666666666,0.7136875)
            (0.3,0.49096875)
            (0.3333333333333333,0.2919375)
            (0.36666666666666664,0.1125625)
            (0.4,0.13978125)
            (0.43333333333333335,0.01621875)
            (0.4666666666666667,0.0165625)
            (0.5,0.0)
        }
        ;
\end{axis}
\end{tikzpicture}
    \caption{\label{fig:remove_head_repulsion}Removing the head repulsion.
    We change the value of the penalty shift function on the blocking part.
    The value $0.5$ equals that on the matching part.
    Failure modes with $0.5$ are shown on the left.
    }
    \vspace{0.2cm}
    \begin{tabular}{|C{2cm}|}
        \hline
        \includegraphics[width=2cm,height=2cm,keepaspectratio,valign=m]{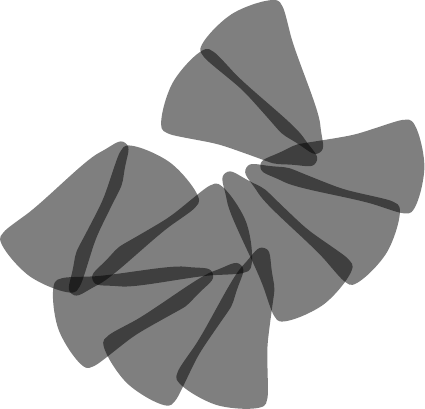} \\
        \hline
        \includegraphics[width=2cm,height=2cm,keepaspectratio,valign=m]{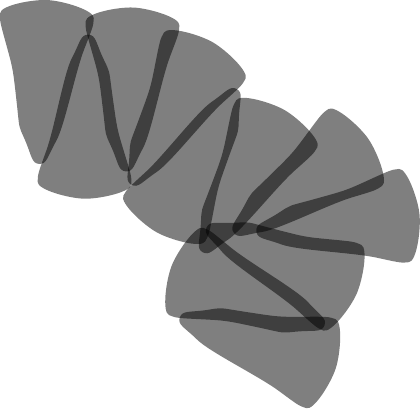} \\
        \hline
    \end{tabular}
    \hspace{0.3cm}
    \begin{tikzpicture}[baseline={(0,0.9cm)}]
\begin{axis}[xlabel={transition to triangle shape},ylabel={self-assembly rate}, width={5cm}, height={4cm}, grid={major}, xmin=0, xmax=1.0, ymin=0, ymax=1, xtick={0.0,0.2,0.4,0.6,0.8,1.0}]
    \addplot[color=red,mark=*]
        coordinates {
            (0.0,0.90084375)
            (0.06666666666666667,0.81778125)
            (0.13333333333333333,0.734)
            (0.2,0.397625)
            (0.26666666666666666,0.074875)
            (0.3333333333333333,0.076875)
            (0.4,0.03190625)
            (0.4666666666666667,0.0524375)
            (0.5333333333333333,0.1286875)
            (0.6,0.121875)
            (0.6666666666666666,0.09934375)
            (0.7333333333333333,0.0865)
            (0.8,0.01796875)
            (0.8666666666666667,0.007625)
            (0.9333333333333333,0.00075)
            (1.0,0.0)
        }
        ;
\end{axis}
\end{tikzpicture}
    \caption{\label{fig:remove_curve}Removing the curvature.
    We transition from the default shape ($0$) to a triangle shape ($1$).
    Failure modes with the triangle shape are shown on the left.
    }
\end{figure}

\begin{figure}
    \begin{tabular}{|C{2cm}|}
        \hline
        \includegraphics[width=2cm,height=2cm,keepaspectratio,valign=m]{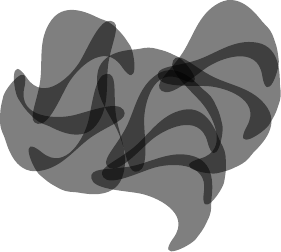} \\
        \hline
        \includegraphics[width=2cm,height=2cm,keepaspectratio,valign=m]{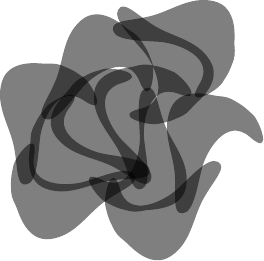} \\
        \hline
    \end{tabular}
    \hspace{0.3cm}
    \begin{tikzpicture}[baseline={(0,0.9cm)}]
\begin{axis}[xlabel={transition to headless shape},ylabel={self-assembly rate}, width={5cm}, height={4cm}, grid={major}, xmin=0, xmax=1.0, ymin=0, ymax=1, xtick={0.0,0.2,0.4,0.6,0.8,1.0}]
    \addplot[color=red,mark=*]
        coordinates {
            (0.0,0.901875)
            (0.06666666666666667,0.9360625)
            (0.13333333333333333,0.92215625)
            (0.2,0.906125)
            (0.26666666666666666,0.91025)
            (0.3333333333333333,0.84925)
            (0.4,0.575625)
            (0.4666666666666667,0.3150625)
            (0.5333333333333333,0.076125)
            (0.6,0.0180625)
            (0.6666666666666666,0.01584375)
            (0.7333333333333333,0.009)
            (0.8,0.0025)
            (0.8666666666666667,0.00253125)
            (0.9333333333333333,0.00078125)
            (1.0,0.00296875)
        }
        ;
\end{axis}
\end{tikzpicture}
    \caption{\label{fig:remove_head}Removing the head.
    We transition from the default shape ($0$) to a headless shape ($1$).
    Failure modes with the headless shape are shown on the left.
    }
\end{figure}
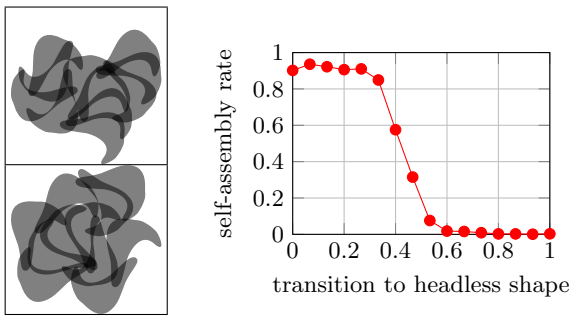

We investigate how the following changes to the shape influence self-assembly robustness:
\begin{itemize}
    \setlength\itemsep{0em}
\item Removing the head repulsion (Fig. \ref{fig:remove_head_repulsion})
\item Removing the curvature (Fig. \ref{fig:remove_curve})
\item Removing the head (Fig. \ref{fig:remove_head}).
\end{itemize}
For each case, we consider a smooth transition from the original shape to the modified version and carry out multiple HMC runs for various points during the transition.
We see in the figures that the self-assembly robustness decays during this transition.

When self-assembly fails, the question is whether this is a failure of the energy function (the energy minimizer no longer corresponds to the assembled state) or a failure of the evolution model (HMC gets stuck in local minima).
In the three cases that we investigate here, the energy minimizer still seems to be the fully assembled configuration.
However, the energy difference between the minimizer and the mis-assembled configurations that HMC gets stuck in is small and the evolution algorithm is unable to find a path to the global minimizer. 


\subsection*{Ground states and symmetry breaking}
We use gradient descent to find stationary points (or ground states) starting from correctly assembly configurations.

\begin{figure}
\centering
\includegraphics[width=3cm]{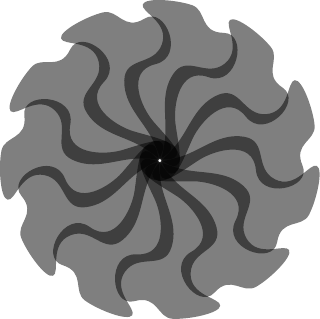}
\hspace{0.1cm}
\begin{tikzpicture}[baseline={(0,-0.3cm)}]
\begin{axis}[xlabel={$i$-th eigenvalue}, width={5cm}, height={4cm}, grid={major}, ymode={log}, xmin=1, xmax=33, ymin={0.01}, ymax={100}, ytick={0.01,0.1,1,10,100}]
    \addplot[only marks,color={red}, mark={*},mark size=1pt]
        coordinates {
            (1,0.010397629744998488)
            (2,0.02297368667939891)
            (3,0.03959077274639617)
            (4,2.364258816038711)
            (5,2.3733862023006473)
            (6,5.2982288213583475)
            (7,5.315043169646526)
            (8,6.5088308602125124)
            (9,6.5196737652643)
            (10,7.546791854713423)
            (11,7.556675511334043)
            (12,9.289910277771762)
            (13,9.787157010761923)
            (14,9.797531338460331)
            (15,14.675834516028482)
            (16,14.703097757997837)
            (17,19.760842751266498)
            (18,26.871846939493622)
            (19,26.91230147349894)
            (20,32.68974071420911)
            (21,32.751291733597185)
            (22,37.94271447222117)
            (23,37.9845384727717)
            (24,46.40505275034849)
            (25,46.45104452601154)
            (26,50.12757037149035)
            (27,50.15136057873855)
            (28,63.80601870882765)
            (29,63.83742617469675)
            (30,72.36060366824739)
            (31,72.4071629671882)
            (32,77.71887774316878)
            (33,77.76048321163421)
        }
        ;
\end{axis}
\end{tikzpicture}
\caption{\label{fig:groundstate_default}Ground state of default shape. Eigenvalues of Hessian shown on the right.}
\end{figure}
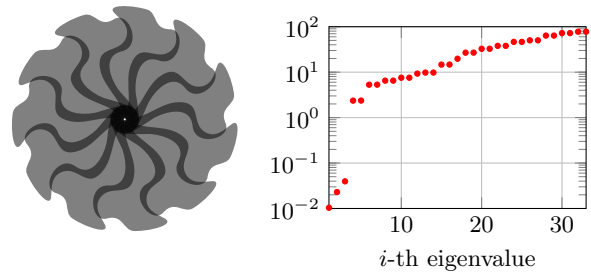

The shape we used up to now leads to a perfectly symmetric ground state.
This state, together with the eigenvalues of its Hessian, is shown in Fig. \ref{fig:groundstate_default}.
The three eigenvalues near zero correspond to the zero modes of translating and rotating the whole configuration.
These eigenvalues are not exactly zero since the energy is not perfectly invariant under these operations due to the use of numerical quadrature on a finite grid.
Since the rest of the eigenvalues are positive, we conclude that the ground state is a local energy minimizer.

\begin{figure}
\centering
\begin{tabular}{cccc}
    \includegraphics[width=.17\textwidth]{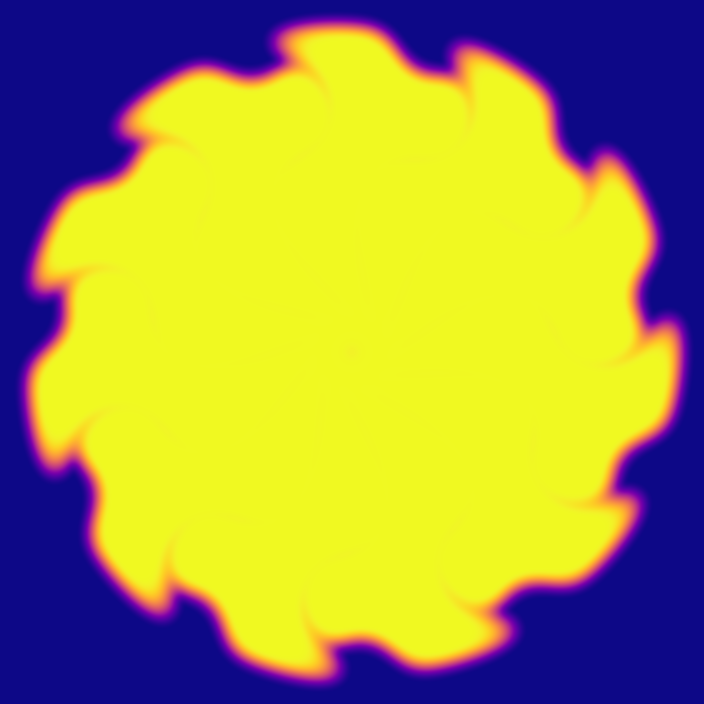} &
    \includegraphics[width=.17\textwidth]{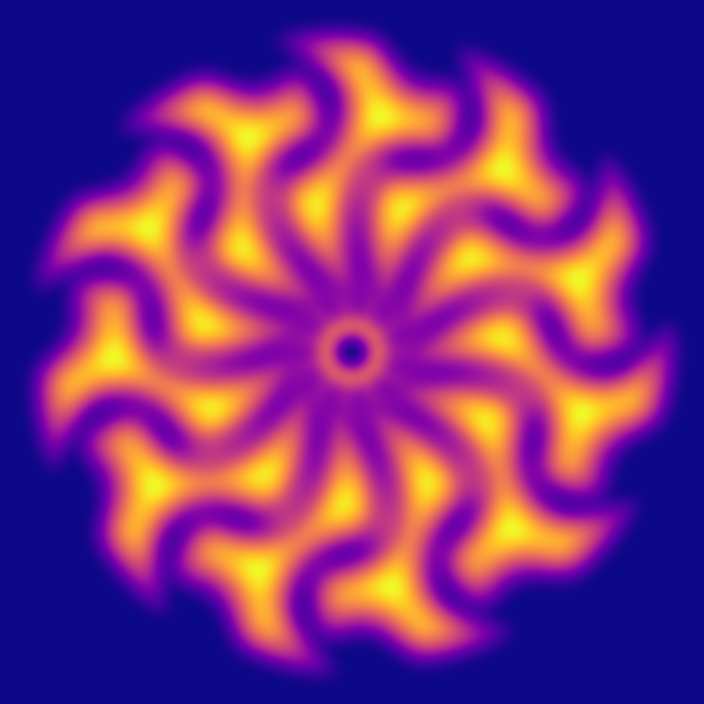} \\
    volume & penalty \\[0.7em]
    \includegraphics[width=.17\textwidth]{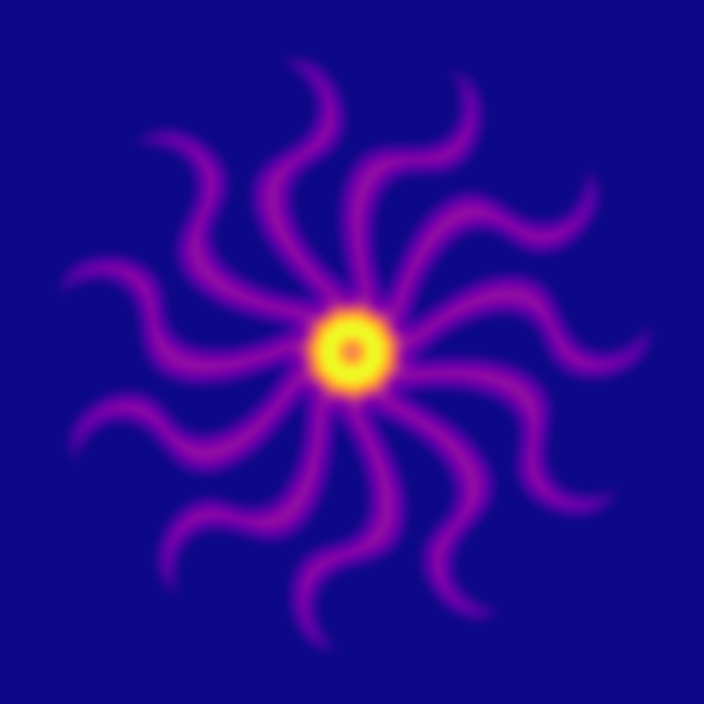} &
    \includegraphics[width=.17\textwidth]{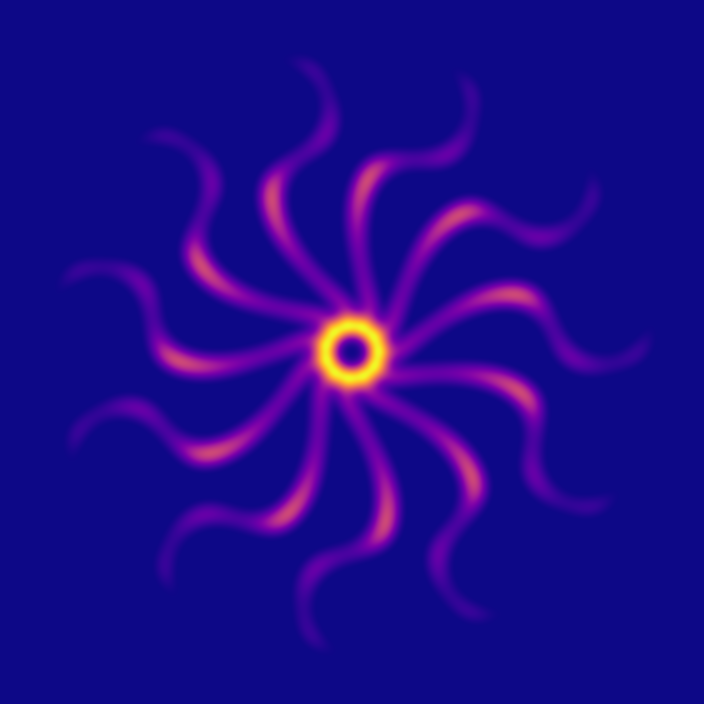} \\
    volume interaction & penalty interaction
\end{tabular}
\caption{\label{fig:gsintegrands}Visualization of the energy integrands. Similar to Fig. \ref{fig:ballintegrands}, but for the ground state shown in Fig. \ref{fig:groundstate_default}.}
\end{figure}

In Fig. \ref{fig:gsintegrands} we show the energy integrands for the ground state.

\begin{figure}
    \centering
    \begin{tabular}{|C{2cm}|}
        \hline
        \includegraphics[width=2cm,height=2cm,keepaspectratio,valign=m]{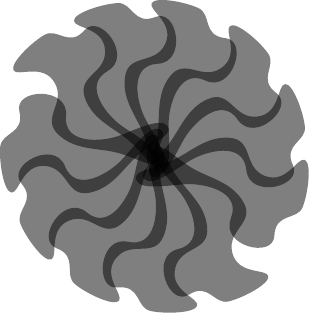} \\
        \hline
        \includegraphics[width=2cm,height=2cm,keepaspectratio,valign=m]{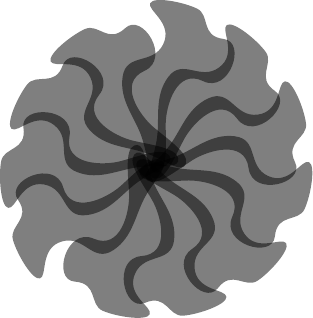} \\
        \hline
    \end{tabular}
    \hspace{0.3cm}
    \begin{tikzpicture}[baseline={(0,0.9cm)}]
\begin{axis}[xlabel={\small shape thinning factor}, ylabel={\small self-assembly rate}, width={5cm}, height={4cm}, grid={major}, xmin=0.95, xmax=1.0, ymin=0, ymax=1, xtick={0.95,0.96,0.97,0.98,0.99,1.0}, ticklabel style = {font=\small}]
    \addplot[color=red,mark=*]
        coordinates {
            (0.95,0.473265625)
            (0.96,0.41953125)
            (0.97,0.3950625)
            (0.98,0.306828125)
            (0.99,0.278859375)
            (1.0,0.2789375)
        }
        ;
\end{axis}
\end{tikzpicture}
    \caption{\label{fig:symbr}A slightly thinner shape self-assembles faster but leads to ground states with broken symmetry.
    Assembly rate computed from HMC iterations $1.5 \cdot 10^3$ to $2 \cdot 10^3$.}
\end{figure}
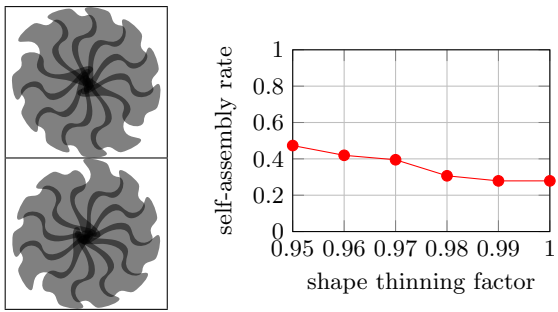

The symmetry of the ground state is sensitive to small modifications of the shape.
To explore this, we rescale our shape along an axis to make it slightly thinner (down to a factor of $0.95$).
Thinning the shape speeds up the self-assembly while also causing symmetry breaking for the ground states (see Fig. \ref{fig:symbr}).
If the shape is thinned even more, an additional $12$th shape would be required for complete assembly.

\section{\label{sec:discussion}Discussion}
We have given a proof-of-concept that demonstrates the potential of our coarse-grained energy and evolution models for efficiently simulating large-scale self-assembly processes of proteins.
Our simulations of nonconvex model shapes indicate the significance of two factors for the robust self-assembly into a unique structure: blocking and matching (i.e., local repulsion and attraction) of different parts of the boundary; and nonconvexity and handedness of the shape. 

Our work can be extended in various ways.
The obvious one is simulating self-assembly in 3D, which we are aiming for with future work.
To simulate protein self-assembly in 3D, the solvent-accessible surface \cite{connolly_solvent-accessible_1983} would be the natural choice for the shape that we would use for our simulations.

Our energy model can be refined into a more accurate model for the free solvation energy using the morphometric approach \cite{hadwiger_integralsatze_1956,hansen-goos_solvation_2007}
\begin{equation}
    E = p\, V + \sigma\, A + \kappa\, C + \tilde{\kappa}\, X,
    \label{eq:generalenergy}
\end{equation}
where the terms are volume ($V$), surface area ($A$), mean width ($C$), and Euler characteristic ($X$).
Our method currently only accounts for the volume term. While this term is at least one order of magnitude bigger than the other terms, the remaining terms provide useful corrections \cite{hansen-goos_density_2006}. 

\begin{acknowledgments}
The research of Lukas Mayrhofer was funded by the Deutsche Forschungsgemeinschaft (DFG, German Research Foundation) via project 195170736 - TRR109.
We thank Ivan Spirandelli for helpful discussions on the modeling and Folkmar Bornemann for suggesting the HMC algorithm.
\end{acknowledgments}

\appendix
\section{\label{sec:sdfapprox}Approximating signed distance functions}
To approximate the signed distance function (SDF) of a shape $S \subset \Rr^2$, we first sample the exact SDF on a fine grid in a sufficiently large box containing $S$.
We reduce aliasing by convolving the result with a Gaussian kernel before subsampling the values on a coarse grid.

From the data on the coarse grid, we construct a smooth SDF approximation using cubic B-spline functions.
In this construction, we do not interpolate.
Instead, we directly take the subsampled values as coefficients for the tensorized B-spline functions.

The SDF approximation procedure adds the following parameters to our energy model:
\begin{itemize}
    \setlength\itemsep{0em}
    \item SDF fine and coarse grid spacing (for sampling the exact signed distance function respectively for B-spline coefficients of approximate signed distance function)
    \item SDF Gaussian standard deviation (standard deviation of Gaussian kernel used to low-pass filter the samples on the fine grid)
\end{itemize}

\section{\label{sec:hmcinvariance}Invariance of Gibbs measure}
To show that the Gibbs measure $\mu$ is invariant under an HMC iteration, we check that the transition kernel $T$ of HMC satisfies detailed balance with respect to $\mu$, that is,
\begin{equation}
\int_B T(x, A) \dif \mu(x) = \int_A T(x, B) \dif \mu(x)
\end{equation}
for all measurable sets $A$ and $B$.
This follows from a straightforward calculation that is not specific to our Lie group setting.
An important ingredient in this calculation is that the Leapfrog algorithm is time-reversible and volume-preserving.

To show these properties, we recall the Leapfrog algorithm:

\algdef{SE}[REPEATN]{RepeatN}{End}[1]{\algorithmicrepeat\ #1 \textbf{times}}{\algorithmicend}
\begin{algorithmic}
  \State $p \gets p - \frac{\varepsilon}{2T} \nabla E(x)$
  \State $x \gets \exp_G(\varepsilon p) x$
  \RepeatN{$L-1$}
  \State $p \gets p - \frac{\varepsilon}{T} \nabla E(x)$
  \State $x \gets \exp_G(\varepsilon p) x$
  \End
  \State $p \gets p - \frac{\varepsilon}{2T} \nabla E(x)$
\end{algorithmic}
Time-reversibility follows since
\begin{equation}
    \exp_G(-p) = \exp_G(p)^{-1}.
\end{equation}

To show that Leapfrog is volume-preserving, we note that it is a composition of maps of the form $(x, p) \mapsto (f(p) x, p)$ and $(x, p) \mapsto (x, p + g(x))$.
The result then follows by applying Fubini's theorem.

\section{\label{sec:shapedefinition}Description of model protein shape}
Our 2D model shape is defined as the region enclosed by a curve that consists of cubic Beziér segments.
Each segment $c \from [0, 1] \to \Rr^2$ is given by
\begin{equation}
    c(t) := (1-t)^3 \beta^0 + 3(1-t)^2 t \beta^1 + 3(1-t) t^2 \beta^2 + t^3 \beta^3,
\end{equation}
where $\beta^0, \beta^1, \beta^2, \beta^3 \in \Rr^2$ are the \emph{control points}.

Table \ref{fig:shapedef} lists the control points for each segment.
The segments were chosen such that the composite curve has a continuous derivative. 
\begin{table}[H]
    \centering
    \includegraphics[width=.23\textwidth]{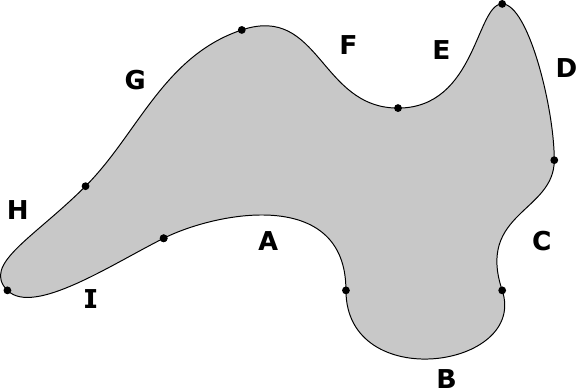}
    \\[1em]
    \bgroup
    \renewcommand{\arraystretch}{1.0}
    $\begin{array}{rrrrrrrrrr}
        \toprule
        & \text{A} & \text{B} & \text{C} & \text{D} & \text{E} & \text{F} & \text{G} & \text{H} & \text{I} \\
        \midrule
        \beta_1^0 & -2 & 5 & 11 & 13 & 11 & 7 & 1 & -5 & -8 \\
        \beta_2^0 & -3 & -5 & -5 & 0 & 6 & 2 & 5 & -1 & -5 \\
        \beta_1^1 & 0 & 5 & 10 & 13 & 10 & 4 & -2 & -7 & -7 \\
        \beta_2^1 & -2 & -9 & -2 & 2 & 6 & 2 & 4 & -3 & -6 \\
        \beta_1^2 & 5 & 12 & 13 & 12 & 10 & 4 & -3 & -9 & -4 \\
        \beta_2^2 & -1 & -8 & -2 & 6 & 2 & 6 & 1 & -4 & -4 \\
        \beta_1^3 & 5 & 11 & 13 & 11 & 7 & 1 & -5 & -8 & -2 \\
        \beta_2^3 & -5 & -5 & 0 & 6 & 2 & 5 & -1 & -5 & -3 \\
        \bottomrule
    \end{array}$
    \egroup
    \caption{\label{fig:shapedef}Description of the model shape}
\end{table}
\section{Simulation parameters}
\begin{table}[H]
    \centering
    \begin{tabular}{p{5.7cm}p{1cm}}
        \toprule
        parameter & value \\
        \midrule
        number of copies of the shape & $11$ \\
        size of ambient space ($\lambda$) & $64$ \\
        smooth step and ramp transition speeds & $(1, 1)$ \\
        penalty shift value on matching part & $0.5$ \\
        penalty shift value on blocking part & $0.1$ \\
        penalty strength & $4$ \\
        quadrature grid spacing & $1$ \\
        SDF fine grid spacing & $0.1$ \\
        SDF coarse grid spacing & $1$ \\
        SDF Gaussian standard deviation & $0.5$ \\
        temperature & $1.5$ \\
        inner product parameters ($\tau_T, \tau_R$) & $(1, 0.21)$ \\
        Leapfrog stepsize & $0.16$ \\
        Leapfrog iterations & $15$ \\
        \bottomrule
    \end{tabular}
    \caption{\label{tab:shapepars}Simulation parameters.}
\end{table}
\bibliography{main}
\end{document}